\documentclass[12pt, a4paper]{article}

\usepackage{amsmath,amssymb,cite,comment,bm,url}

\input{colordvi.tex}

\usepackage{ifpdf}
\ifpdf        
\usepackage{graphicx, hyperref, xcolor}     
\else     
\usepackage[dvipdfmx]{graphicx, hyperref, xcolor}     
\fi

\usepackage[height=21.5cm,width=16.5cm]{geometry}

\definecolor{rossoferrari}{HTML}{D9073D}
\definecolor{mediumblue}{HTML}{0000CD}
\hypersetup{
setpagesize=false,
bookmarksnumbered=true,
bookmarksopen=true,
colorlinks=true,
linkcolor=rossoferrari,
urlcolor=mediumblue,
citecolor=mediumblue,
}

\begin{document}

\begin{titlepage}

\begin{center}

\hfill DESY 20-104\\

\vskip 2cm

{\Large \bf 
High-frequency Graviton from Inflaton Oscillation
}

\vskip 1cm

{\large
Yohei Ema$^{(a)}$, Ryusuke Jinno$^{(a)}$ and Kazunori Nakayama$^{(b,c)}$
}

\vskip 1cm

$^{(a)}${\em DESY, Notkestrabe 85, D-22607 Hamburg, Germany}\\[.3em]
$^{(b)}${\em Department of Physics, Faculty of Science,\\
The University of Tokyo,  Bunkyo-ku, Tokyo 113-0033, Japan}\\[.3em]
$^{(c)}${\em Kavli IPMU (WPI), The University of Tokyo,  Kashiwa, Chiba 277-8583, Japan}

\end{center}
\vskip 1cm

\begin{abstract}

We point out that there is a high-frequency tail of the stochastic inflationary gravitational wave background that scales as $f^{-1/2}$ with frequency $f$. This contribution comes from the graviton vacuum fluctuation amplified by the inflaton coherent oscillation during the reheating stage.
It contains information on inflaton properties such as the inflaton mass as well as the thermal history of the early Universe.

\end{abstract}

\end{titlepage}


\renewcommand{\thepage}{\arabic{page}}
\setcounter{page}{1}
\renewcommand{\thefootnote}{\#\arabic{footnote}}
\setcounter{footnote}{0}

\newpage

\tableofcontents

\section{Introduction}
\label{sec:Intro}

Gravitational wave (GW) provides us with a new way to probe our universe. It interacts only very weakly with matters, and hence preserves the information on source objects imprinted in its spectrum during its propagation.
In particular, GW is a unique way to probe the early universe, as all the other messengers (such as photons and neutrinos) interact strongly with matters and hence lost their information in the (sufficiently) early universe.
For instance, inflation generically predicts GWs that are excited during the quasi de Sitter phase~\cite{Starobinsky:1979ty,Maggiore:2018sht,Giovannini:2019oii}.
They are one of the main targets of the modern cosmological observatories since they provide us with the information on the inflationary energy scale.
GWs are also expected to be produced during the course of thermal history after inflation, such as preheating~\cite{Khlebnikov:1997di}, cosmological defects~\cite{Vilenkin:2000jqa}, and phase transitions~\cite{Witten:1984rs,Hogan:1986qda}, although the existence of these contributions is more model-dependent (see Ref.~\cite{Caprini:2018mtu} for a recent review).

In this paper, we point out the existence of yet another source of GW. In general, an inflaton starts to oscillate around the bottom of its potential after inflation, before eventually decaying into other particles and hence completing the reheating.
This coherent oscillation of the inflaton during the inflaton oscillation epoch produces GWs through gravitational interaction, which can be interpreted as inflaton annihilation into gravitons mediated by gravity itself~\cite{Ema:2015dka}.
This contribution extends toward the high frequency region beyond the inflationary GWs.
Since it is produced from the inflaton oscillation, the GW spectrum contains a variety of information on the inflaton sector, such as the inflaton mass scale and more generally the shape of the inflaton potential around its bottom.
It also depends on the inflaton decay rate and hence the reheating temperature.

It is quite challenging to detect this contribution with the current and near-future GW detectors~\cite{Kawamura:2006up,Punturo:2010zz,Janssen:2014dka,TheLIGOScientific:2016dpb,amaroseoane2017laser,Graham:2017pmn,Bertoldi:2019tck} (see also Refs.~\cite{Li:2008qr,Li:2009zzy,Ejlli:2019bqj,Ito:2019wcb,Ito:2020wxi} for ideas for high-frequency graviton detection), not only because of its overall normalization but also because of its high characteristic frequency.
Furthermore, there are other GW sources in the high frequency region that we expect to be present in general, such as the contribution from the standard model (SM) thermal plasma~\cite{Ghiglieri:2015nfa,Ghiglieri:2020mhm}, the bremsstrahlung from the inflaton decay~\cite{Nakayama:2018ptw,Huang:2019lgd} as well as those from preheating~\cite{Khlebnikov:1997di}.
These other contributions can hide our GWs, depending on the model parameters and the frequency.
Nevertheless, we think it meaningful to point out the existence of the GWs produced during the inflaton oscillation epoch, as this contribution imprints quite interesting information on the inflaton sector that is usually hard to reach.
We hope that a development of GW detection technology eventually enables us to probe the high frequency region such that we can gain information on the inflaton sector in the future.

This paper is organized as follows.
In Sec.~\ref{sec:grav}, we review the equation of motion of the graviton and the graviton production during inflation.
Sec.~\ref{sec:high} is the main part of this paper, where we compute the GW production from the inflaton oscillation after inflation both analytically and numerically.
In Sec.~\ref{sec:stochasticGW}, we show the resultant GW spectrum, especially its dependence on the model parameters such as the inflaton mass and the reheating temperature.
Sec.~\ref{sec:discussion} is devoted to the discussion, where we compare our contribution with the contributions from the SM thermal plasma and the bremsstrahlung from the inflaton decay.

\section{Graviton in inflationary universe}
\label{sec:grav}

We consider the Einstein-Hilbert action plus the inflaton action as
\begin{align}
	S=\int dtd^3x \sqrt{-g} \left( \frac{M_P^2}{2}R + \mathcal L_\phi \right).
\end{align}
The metric is expanded as
\begin{align}
	ds^2= -dt^2 + a^2(t)(\delta_{ij}+h_{ij})dx^idx^j = a^2(t)\left[-d\tau^2+ (\delta_{ij}+h_{ij})dx^idx^j \right],
\end{align}
where we have taken the transverse-traceless gauge: ${h^i}_i=\partial^ih_{ij}=0$. The graviton action is given by
\begin{align}
	S &= \int dtd^3x\,a^3\frac{M_P^2}{8}\left[ (\dot h_{ij})^2-\frac{1}{a^2}(\partial_l h_{ij})^2\right] \nonumber \\
	&=\sum_{\lambda=+,\times}\int \frac{d\tau d^3k}{(2\pi)^3} \frac{1}{2}\left[ 
	\left| h'_\lambda(k)\right|^2- \omega_k^2\left| h_\lambda(k)\right|^2\right],~~~~~~
	\omega_k^2\equiv k^2-\frac{a^2 R}{6},
	\label{S_grav}
\end{align}
where the prime denotes the derivative with respect to the conformal time $\tau$, and $\lambda=+,\times$ denotes the two polarization states of the graviton. We have defined the canonical graviton in the momentum space as
\begin{align}
	\frac{aM_P}{2}h_{ij}(t,\vec x) = \sum_{\lambda=+,\times}\int\frac{d^3k}{(2\pi)^3} h_\lambda(\vec k,\tau) e^{i\vec k\cdot \vec x} \epsilon_{ij}^\lambda,
\end{align}
where $\epsilon^\lambda_{ij}$ denotes the polarization tensor, which satisfies $\epsilon_{ij}^{\lambda} \epsilon_{ij}^{\lambda'}=\delta_{\lambda\lambda'}$.
The free graviton action (\ref{S_grav}) is the same as the minimally-coupled massless scalar field. Thus gravitational production of gravitons during the inflation and reheating era is treated in the same way as the minimal scalar field which is extensively studied in Refs.~\cite{Ema:2015dka,Ema:2016hlw,Ema:2018ucl,Chung:2018ayg}. 

Let us introduce a creation and annihilation operator for the graviton:
\begin{align}
	h_\lambda(\vec k,\tau) = \widetilde h_{\lambda}(\vec k,\tau) a_{\lambda,\vec k} +  \widetilde h^*_{\lambda}(\vec k,\tau) a^\dagger_{\lambda,-\vec k},
\end{align}
where they satisfy the commutation relation $\left[ a_{\lambda,\vec k}, a^\dagger_{\lambda',\vec k'}\right]=(2\pi)^3\delta(\vec k-\vec k')\delta_{\lambda\lambda'}$. The equation of motion is given by
\begin{align}
	\widetilde h_\lambda''(k) + \omega_k^2 \widetilde h_\lambda(k) = 0.
\end{align}
The solution to the equation of motion and its approximate form in the high and low frequency limit, which satisfies the Bunch-Davies boundary condition, during inflation is given by
\begin{align}
	\widetilde h_\lambda(k,\tau) = -\frac{1}{\sqrt{2k}} \sqrt{\frac{-\pi k\tau}{2}} H_{3/2}^{(1)}(-k\tau)
	\simeq \begin{cases}
	\displaystyle \frac{1}{\sqrt{2k}} e^{-ik\tau} & {\rm for}~~ -k\tau\gg 1\\
	\displaystyle i \frac{a H_{\rm inf}}{\sqrt{2} k^{3/2}} & {\rm for}~~ -k\tau\ll 1
	\end{cases},
\end{align}
where $H_{3/2}^{(1)}(x)$ denotes the Hankel function of the first kind and we used $\tau=-(aH_{\rm inf})^{-1}$ during inflation with $H_{\rm inf}$ being the inflationary Hubble scale.
It is well known that the superhorizon modes ($-k\tau_{\rm end} \lesssim 1$ where the subscript ``end'' represents the end of inflation) have (nearly) scale invariant power spectrum:\footnote{
	Often the graviton power spectrum is defined by the original basis before the canonical rescaling. In such a case the graviton power spectrum is given by $\overline{\mathcal P}_h(k) \equiv (2/M_P)^2\mathcal P_h(k) = 2H_{\rm inf}^2/(\pi M_P)^2$ and the tensor-to-scalar ratio is defined as $r=\overline{\mathcal P}_h(k)/\mathcal P_\zeta(k)$ with $\mathcal P_\zeta$ being the power spectrum of the curvature perturbation.
}
\begin{align}
	\mathcal P_h(k,\tau_{\rm end}) \equiv \frac{k^3}{\pi^2a^2} \left|\widetilde h_\lambda(k)\right |^2= \frac{H_{\rm inf}^2}{2\pi^2}~~~~~~{\rm for}~~~-k\tau_{\rm end}\ll 1.
\end{align}
On the other hand, shorter wavelength modes $(-k\tau_{\rm end}\gtrsim 1)$ never exit the horizon. However, it does not mean that shorter wavelength modes are not excited. Below we will evaluate the production of 
these short wavelength graviton modes during the inflaton oscillation epoch after inflation.

\section{High frequency graviton production}
\label{sec:high}

Let us consider the high-frequency modes that never exit the horizon: $-k\tau_{\rm end}\gtrsim 1$. After inflation ends, the inflaton coherent oscillation begins and the graviton wave function is modified through the (rapidly-oscillating) $a^2 R$ term in the equation of motion. In this case it is convenient to parameterize the wave function in terms of the Bogoliubov coefficients $\alpha_k, \beta_k$ as
\begin{align}
	\widetilde h_\lambda(k,\tau) = \alpha_k(\tau)v_k(\tau) + \beta_k(\tau) v_k^*(\tau),
\end{align}
where
\begin{align}
	v_k(\tau) = e^{-i\Omega_k(\tau)},~~~\Omega_k(\tau) \equiv \int\omega_k d\tau.
\end{align}
The equation of motion is rewritten as
\begin{align}
	\alpha_k'(\tau) = \frac{\omega_k'}{2\omega_k}\beta_k(\tau)e^{2i\Omega_k},~~~~~~
	\beta_k'(\tau) = \frac{\omega_k'}{2\omega_k}\alpha_k(\tau)e^{-2i\Omega_k},
	\label{beta}
\end{align}
They satisfy the normalization condition $|\alpha_k(\tau)|^2-|\beta_k(\tau)|^2=1$. The initial condition is $\alpha_k=1$ and $\beta_k=0$ for $-k\tau\to \infty$. The renormalized graviton energy density is expressed as
\begin{align}
	a^4(\tau) \rho_h(\tau) = 2\int\frac{d^3k}{(2\pi)^3} \omega_k \left| \beta_k(\tau) \right|^2.
\end{align}
We define the graviton energy spectrum as
\begin{align}
	\rho_h(\tau)=\int\rho_{h,k}(\tau)d\ln k,~~~~a^4(\tau)\rho_{h,k}(\tau) = \frac{k^3\omega_k}{\pi^2}\left| \beta_k(\tau) \right|^2.
\end{align}
Thus it is sufficient to evaluate $\beta_k(\tau)$ to obtain the graviton energy spectrum. Numerically, one can integrate the equation (\ref{beta}) to obtain $\beta_k(\tau)$ given an inflation model. 

\begin{figure}
\centering
\begin{tabular}{cc}
\includegraphics[width=0.49\columnwidth]{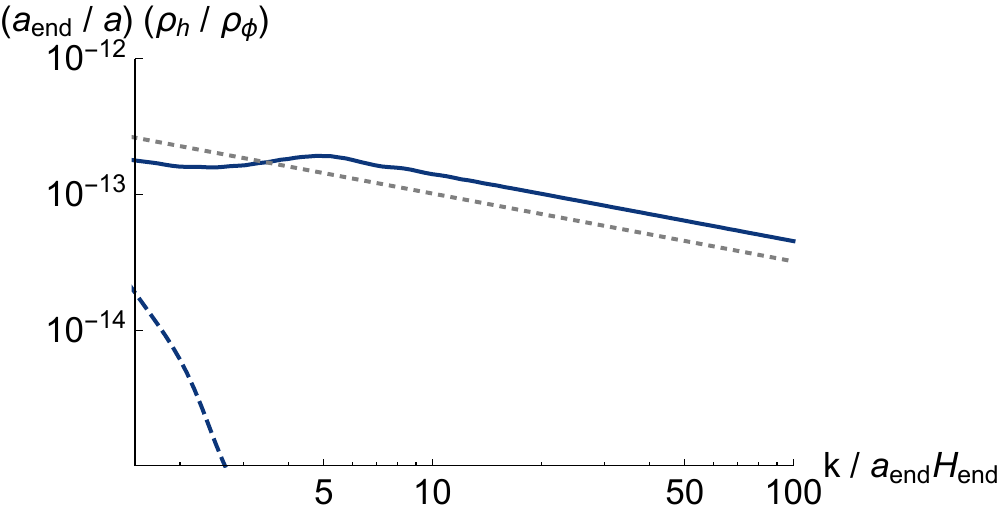}
\includegraphics[width=0.49\columnwidth]{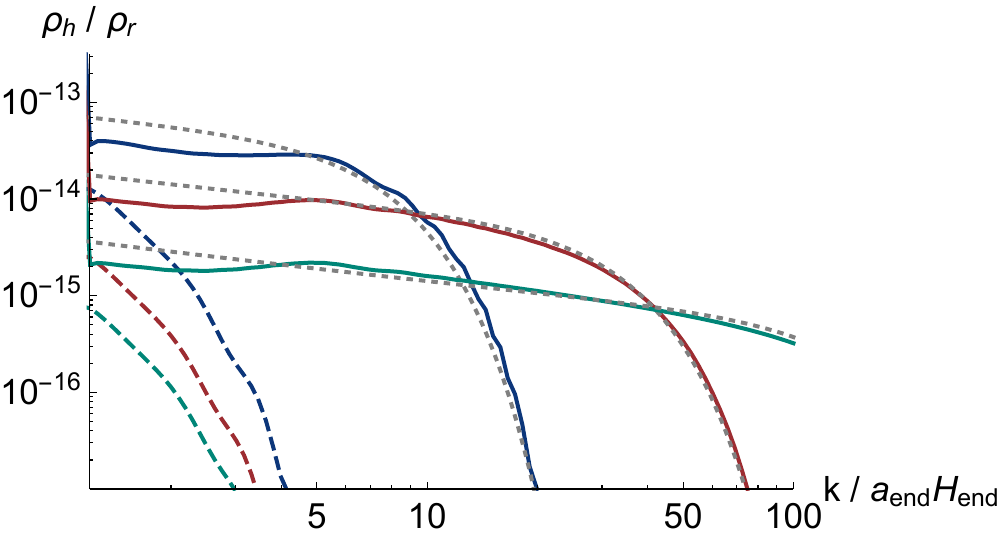}
\end{tabular}
\caption{\small
Graviton energy spectrum after inflation. 
The numerical solutions of Eq.~(\ref{beta}) with Eqs.~\eqref{eq:Friedmann}-\eqref{eq:rhor} are shown as the solid lines, 
while the analytic estimates, \textit{i.e.} Eq.~\eqref{rhok_ana}, are shown as the gray-dotted line,
for the quadratic inflaton potential $V=m_\phi^2\phi^2/2$. 
For comparison, the numerical results without the inflaton oscillation (Eq.~\eqref{eq:rhophi})
are shown as the dashed line. 
The modes shown here never exit the horizon. 
(Left) The inflaton decay rate is taken to vanish, and the spectrum is evaluated during the inflaton domination. 
The ratio $\rho_h / \rho_\phi$ is multiplied with the ratio of the scale factor at the inflation end and at the evaluation time to cancel out the dependence on the evaluation time.
(Right) The inflaton decay rate is taken to be $\Gamma_\phi=10^{-1} m_\phi$ (blue), $10^{-2} m_\phi$ (red) and $10^{-3} m_\phi$ (green), and the spectrum is evaluated during the radiation domination.
}
\label{fig:grav}
\end{figure}

Fig.~\ref{fig:grav} shows the result of our numerical calculation. We assumed a chaotic inflation model with a quadratic potential $V=m_\phi^2\phi^2/2$ for concreteness,\footnote{The chaotic inflation~\cite{Linde:1983gd} with a quadratic potential is now disfavored by the cosmological observation, but a slight modification on the potential makes the model viable~\cite{Destri:2007pv,Nakayama:2013jka}. Since we are mainly interested in the inflaton oscillation regime, such a modification is irrelevant for the discussion below.} and solved the following equations
\begin{align}
&3 M_P^2 {\cal H}^2 = a^2 (\rho_\phi + \rho_r),
\label{eq:Friedmann}
\\
&\phi'' + (2 {\cal H} + a \Gamma_\phi) \phi' + a^2 \frac{dV}{d\phi} = 0,
\label{eq:phi}
\\
&\rho_r' + 4 {\cal H} \rho_r = a \Gamma_\phi \rho_\phi,
\label{eq:rhor}
\end{align}
where $\rho_\phi = \phi'^2/2a^2 + V$ is the inflaton energy density, $\rho_r$ is the radiation energy density, $\Gamma_\phi$ is the inflaton total decay width and $\mathcal H=a'/a$ denotes the conformal Hubble scale.
Eq.~(\ref{beta}) is solved numerically with this background. The inflaton decay rate is (hypothetically) taken to be zero in the left panel and the spectrum is evaluated during the inflaton domination. 
In the right panel the inflaton decay rate is taken to be $\Gamma_\phi=(10^{-1}, 10^{-2}, 10^{-3}) \times m_\phi$ and the spectrum is evaluated during the radiation domination.
One can see that the graviton energy spectrum shows $k^{-1/2}$ behavior as expected. Below we compare this result with an analytic estimate.

In order to evaluate the graviton energy spectrum, it is convenient to interpret the graviton production during the reheating era as the inflaton annihilation into a graviton pair, as emphasized in Refs.~\cite{Ema:2015dka,Ema:2016hlw,Ema:2018ucl}. Taking account of two polarization states of the graviton, the effective inflaton annihilation rate is estimated as\footnote{
	Ref.~\cite{Chung:2018ayg} calculated the gravitational production rate analytically including $\mathcal O(1)$ numerical factor for a scalar particle. The same result is applied for a graviton production, since the graviton action is the same as the minimal massless scalar.
}
\begin{align}
	\Gamma^{\rm (grav)}(\phi\phi\to hh) \simeq \frac{1}{192\pi}\frac{\rho_\phi m_\phi}{M_P^4}.
\end{align}
Each annihilation produces a pair of gravitons with energy $m_\phi$ that are then redshifted away. Such a process continues until the end of reheating, resulting in a continuum graviton spectrum at the present universe. The graviton spectrum is calculated as
\begin{align}
	\rho_{h,k}(t_0)&={\cal C} \times \rho_\phi(t_k)\frac{\Gamma^{\rm (grav)}(\phi\phi\to hh)}{H_k} \left( \frac{a(t_k)}{a_0} \right)^4\\
	&\simeq\begin{cases} 
	\displaystyle\frac{3{\cal C}}{64\pi}m_\phi H_{\rm end}^3\left( \frac{m_\phi a_{\rm end}}{k} \right)^{1/2} \left( \frac{a_{\rm end}}{a_0} \right)^{4}
	& {\rm for}~~k \lesssim k_{\rm high}, \\
	\displaystyle\frac{3{\cal C}}{64\pi}m_\phi H_{\rm end}^3\left( \frac{m_\phi a_{\rm end}}{k_{\rm high}} \right)^{1/2} \left( \frac{a_{\rm end}}{a_0} \right)^{4}
	e^{-2k^2/k_{\rm high}^2}& {\rm for}~~k \gtrsim k_{\rm high},
	\end{cases}
	\label{rhok_ana}
\end{align}
where ${\cal C}$ is an ${\cal O}(1)$ coefficient and $t_k$ is defined through $k=a(t_k) m_\phi$, i.e., the cosmic time at which the present graviton frequency $k/a_0$ was emitted. 
Here $k_{\rm high}$ is defined as $k_{\rm high}={\cal C}_k m_\phi a(H=\Gamma_\phi)$ with ${\cal C}_k$ being an ${\cal O}(1)$ coefficient.
The analytic estimate (\ref{rhok_ana}) with ${\cal C} = 3$ and ${\cal C}_k = 1.5$ is also plotted in Fig.~\ref{fig:grav} and it agrees well with the numerical result. 
We used the function $g$ in Eq.~(\ref{g}) to interpolate between $k \lesssim k_{\rm high}$ and $k \gtrsim k_{\rm high}$.
Note that there is a small deviation around $k\sim a_{\rm end} H_{\rm end}$. This is because there is an intermediate epoch around the end of inflation, in which the inflaton oscillation may not be regarded as a harmonic oscillation, while the analytic estimate (\ref{rhok_ana}) assumes the harmonic inflaton oscillation.
For low scale inflation models with a large hierarchy between $m_\phi$ and $H_{\rm inf}$, it is more difficult to treat this intermediate epoch, but the high frequency behavior $k\gtrsim a_{\rm end} m_\phi$ is expected to be well described by the above picture.

In the above picture, the typical momentum of gravitons being produced is constant in time and is around the inflaton mass $m_\phi$. Therefore, it is crucial to use the inflaton equation of motion (\ref{eq:phi}) to get the correct result, since otherwise the scale $m_\phi$ never appears in the system.
In order to stress this point, in Fig.~\ref{fig:grav}, we also show as the dashed lines
the graviton spectrum computed assuming a smooth background evolution
\begin{align}
\rho_\phi
&=
\frac{\rho_{\phi, {\rm inf}}}{1 + (a/a_{\rm end})^3 e^{\Gamma_\phi (t - t_{\rm end})}}
\simeq
\begin{cases} 
\rho_{\phi, {\rm inf}} & {\rm for}~~t < t_{\rm end}, \\[0.2cm]
\displaystyle \rho_{\phi, {\rm inf}} \left( \frac{a}{a_{\rm end}} \right)^{-3} e^{-\Gamma_\phi (t - t_{\rm end})} & {\rm for}~~t > t_{\rm end},
\end{cases}
\label{eq:rhophi}
\end{align}
together with Eqs.~(\ref{eq:Friedmann}) and (\ref{eq:rhor}),
which do not have the timescale $m_\phi$.
One can clearly see that the resultant graviton spectrum is highly suppressed in high frequencies
compared to the solid lines.
Thus, the inflaton oscillation is crucial for the high-frequency behavior of the spectrum.

\section{Stochastic gravitational wave background revisited}
\label{sec:stochasticGW}

Now we plot the present stochastic GW background spectrum in terms of $\Omega_{\rm GW}(k)\equiv \rho_{h,k}/ \rho_{\rm cr}$. 
First, the energy spectrum of subhorizon modes induced by the inflaton oscillation is given by Eq.~(\ref{rhok_ana}) and hence
\begin{align}
	\Omega^{\rm (osc)}_{\rm GW}(k)\simeq 6\times 10^{-24}\left( \frac{m_\phi}{10^{13}\,{\rm GeV}} \right)
	\left( \frac{H_{\rm end}}{10^{13}\,{\rm GeV}} \right)^{1/3}
	\left( \frac{T_{\rm R}}{10^{10}\,{\rm GeV}} \right)^{4/3}
	\left( \frac{a_{\rm end}m_\phi}{k} \right)^{1/2} g(k),
\end{align}
where we used ${\cal C} \simeq 3$ and
\begin{align}
	g(k) \simeq \left[1+ \left(\frac{k}{k_{\rm high}}\right)^{1/2} \right] e^{-2k^2/k_{\rm high}^2}.
	\label{g}
\end{align}
The low and high frequency end of the spectrum are given respectively as
\begin{align}
	f_{\rm low} = \frac{m_\phi}{2\pi}\frac{a_{\rm end}}{a_0}
	 \simeq 1.1\times 10^6\,{\rm Hz} \left( \frac{m_\phi}{10^{13}\,{\rm GeV}} \right)\left( \frac{T_{\rm R}}{10^{10}\,{\rm GeV}}
	 \right)^{1/3}
	 \left( \frac{10^{13}\,{\rm GeV}}{H_{\rm end}} \right)^{2/3},
\end{align}
and
\begin{align}
	f_{\rm high} = \frac{k_{\rm high}}{2\pi a_0}
	\simeq 2.9\times 10^{13}\,{\rm Hz} \left( \frac{m_\phi}{10^{13}\,{\rm GeV}} \right)\left( \frac{10^{10}\,{\rm GeV}}{T_{\rm R}} \right),
\end{align}
where we used ${\cal C}_k \simeq 1.5$. For $f>f_{\rm high}$ the spectrum decays exponentially.
Remember that the present frequency $f$ is related to the comoving wavenumber $k$ through $f=k/(2\pi a_0)$.

On the other hand, there are also contributions from the superhorizon modes that exit the horizon during inflation and reenter the horizon after inflation~\cite{Maggiore:1999vm,Smith:2005mm,Boyle:2005se}. 
The shape of present GW spectrum depends on the equation of state of the Universe. In particular, the GW spectrum scales as $\Omega_{\rm GW} \propto k^0$ $(k^{-2})$ for modes that enter the horizon during radiation (matter) domination~\cite{Nakayama:2008ip,Nakayama:2008wy,Kuroyanagi:2008ye}.
The GW spectrum is evaluated as
\begin{align}
	\Omega^{\rm (inf)}_{\rm GW}(k) \simeq  \Omega_m^2\frac{3r}{128}
	\mathcal P_\zeta(k_0) 
	\left(\frac{k_0}{k}\right)^{2-n_t}
	\left( \frac{g_*(T_k)}{g_*(T_{\rm eq})} \right)
	\left( \frac{g_{*s}(T_{\rm eq})}{g_{*s}(T_k)} \right)^{4/3}
	T_1\left( \frac{k}{k_{\rm eq}}\right) T_2\left( \frac{k}{k_{\rm R}}\right),
\end{align}
where $k_0/a_0=H_0$ is the Hubble parameter at present, $r$ is the tensor-to-scalar ratio, $n_t=-r/8$ is the tensor spectral index, $T_1(x)\simeq 1+(32/9)x^2$ and $T_2(x)\simeq (1+x^2)^{-1}$, and
\begin{align}
	f_{\rm R}=\frac{H_{\rm R}}{2\pi}\frac{a_{\rm R}}{a_0} \simeq 2.6\times 10^2\,{\rm Hz}
	\left( \frac{g_{*s}(T_{\rm R})}{106.75} \right)^{1/6} \left( \frac{T_{\rm R}}{10^{10}\,{\rm GeV}} \right).
\end{align}
This GW spectrum is cut at the frequency $f_{\rm end}$:
\begin{align}
	f_{\rm end} =  \frac{H_{\rm end}}{2\pi}\frac{a_{\rm end}}{a_0}
	\simeq 1.1\times 10^6\,{\rm Hz} \left( \frac{H_{\rm end}}{10^{13}\,{\rm GeV}} \right)^{1/3}\left( \frac{T_{\rm R}}{10^{10}\,{\rm GeV}} \right)^{1/3}.
\end{align}

\begin{figure}
\centering
\begin{tabular}{cc}
\includegraphics[width=0.49\columnwidth]{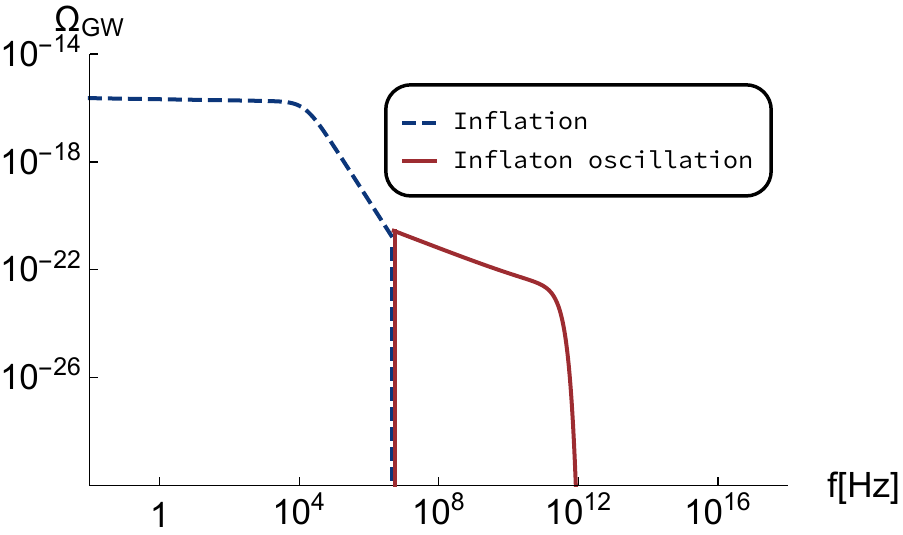}
\includegraphics[width=0.49\columnwidth]{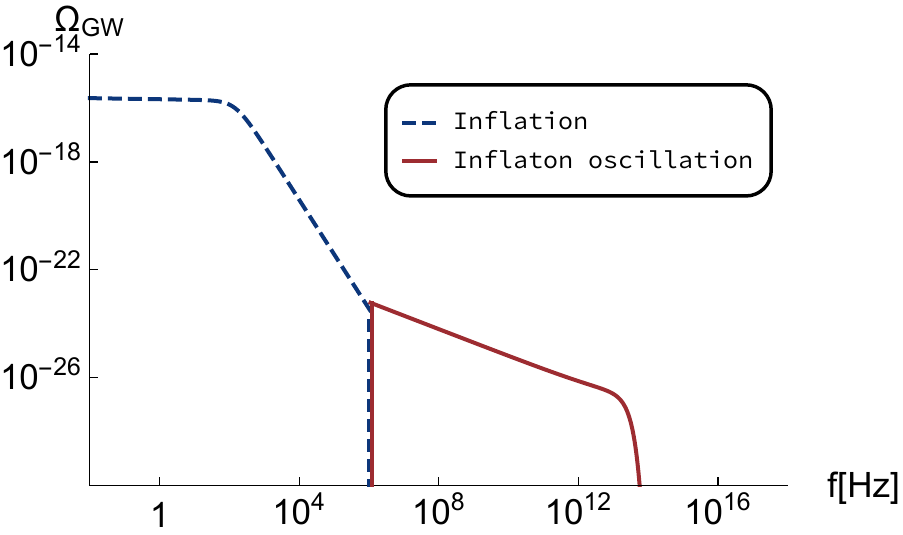}
\end{tabular}
\caption{\small
	Stochastic GW background spectrum for $H_{\rm inf}=10^{14}\,{\rm GeV}, H_{\rm end}=m_\phi=10^{13}$\,GeV, and $T_{\rm R}=10^{12}\,{\rm GeV}$ (left) and $10^{10}\,{\rm GeV}$ (right). The solid lines correspond to the vacuum contribution that is amplified due to the inflaton oscillation during the reheating era and the dashed lines correspond to the inflationary GW that is amplified during the inflation stage.
}
\label{fig:gw}
\end{figure}

Fig.~\ref{fig:gw} shows the stochastic GW background spectrum for $H_{\rm inf}=10^{14}\,{\rm GeV}, H_{\rm end}=m_\phi=10^{13}$\,GeV and $T_{\rm R}=10^{12}\,{\rm GeV}$ (left) and $10^{10}\,{\rm GeV}$ (right). The solid lines correspond to the vacuum contribution that is amplified due to the inflaton oscillation during the reheating era and the dashed lines correspond to the inflationary GW that is amplified during the inflation stage.

\section{Discussion}
\label{sec:discussion}

We have shown that there is inevitably a contribution to the stochastic GW background from the reheating era. It is the subhorizon graviton excitation amplified by the inflaton oscillation. As shown in Fig.~\ref{fig:gw}, this extends to the high-frequency tail which scales as $\Omega_{\rm GW} \propto f^{-1/2}$ in addition to the well-known inflationary GWs that exit the horizon during inflation and reenter the horizon after inflation.
This high frequency tail contains a lot of information about the property of the inflaton: the inflaton mass, the inflaton lifetime (or the reheating temperature) and so on. Although we have focused on the simple quadratic inflaton potential, it is expected that the high frequency tail exhibits more nontrivial structure for a more general form of the inflaton potential.
We will come back to this issue in a separate work.

Lastly we discuss other contributions to the high-frequency stochastic GW background spectrum, which can hide the vacuum contributions that we found.
The Standard Model thermal plasma emits gravitons through scattering processes and they also constitute a stochastic GW background~\cite{Ghiglieri:2015nfa,Ghiglieri:2020mhm}. The typical frequency of the emitted graviton at the temperature $T$ is of order $T$ and it is redshifted as $a^{-1}(t)$. Since the temperature is also redshifted as $a^{-1}(t)$, the typical comoving frequency (or the frequency observed today) is roughly the same independently of the temperature. The overall amount of GW is dominated by those emitted earlier epoch for all the frequency range, {\it i.e.}, at the highest temperature $T_{\rm R}$ and the result is\footnote{
	The dilute plasma before the completion of the reheating also emit gravitons. However, one can show that this contribution goes like $k^{4.6}$ toward lower frequency and is hidden by the $k^3$ tail of (\ref{O_th}).
}
\begin{align}
	\Omega^{\rm (th)}_{\rm GW}(k) 
	\sim 2\times 10^{-13}  \left( \frac{T_{\rm R}}{10^{10}\,{\rm GeV}} \right)\left(\frac{k}{a_*T_*}\right)^3 \varphi\left(\frac{k}{a_*T_*}\right),
	\label{O_th}
\end{align}
where $T_*$ denotes the reference temperature taken to be the electroweak scale, and $\varphi(x)\simeq 1$ for $x \ll 1$ and exponentially decreases for $x\gtrsim 1$.
Another contribution comes from the graviton bremsstrahlung processes associated with the perturbative inflaton decay. The spectrum is given by~\cite{Nakayama:2018ptw,Huang:2019lgd}
\begin{align}
	\Omega^{\rm (brem)}_{\rm GW}(k) \simeq \Omega_{r} \frac{m_\phi^2}{16\pi^2 M_P^2} \frac{f}{f_{\rm high}},
	\label{OGW_brems}
\end{align}
for $f\ll f_{\rm high}$. 
Note that the coupling that is responsible for the inflaton decay may also induce the preheating and resonant particle production if the coupling is relatively large~\cite{Dolgov:1989us,Traschen:1990sw,Kofman:1994rk,Shtanov:1994ce,Kofman:1997yn}. It may act as a classical source of GWs resulting in more abundant GW background than Eq.~\eqref{OGW_brems}~\cite{Khlebnikov:1997di}, but it is rather model dependent and we do not go into details here.
Although in the most parameter regions these contributions are larger than those from the inflaton oscillation, it might be possible to remove these contributions from the data and find the inflaton oscillation signal, which will provide us with rich information on the early universe and the nature of the inflaton.
We emphasize that the high frequency GWs induced by the inflaton oscillation may not be regarded as classical waves since they never exit the horizon and the occupation number is much smaller than unity. Thus detection of such high frequency GWs may be regarded as a direct test of quantum nature of the graviton.

\section*{Acknowledgments}

This work was supported by the Grant-in-Aid for Scientific Research C (No.18K03609 [KN]) and Innovative Areas (No.17H06359 [KN]).
The work of RJ is supported by Grants-in-Aid for JSPS Overseas Research Fellow (No. 201960698). 
This work was supported by the Deutsche Forschungsgemeinschaft under Germany's Excellence Strategy - EXC 2121 ``Quantum Universe" - 390833306. This work was also supported
by the ERC Starting Grant ``NewAve" (638528).

\small
\bibliographystyle{utphys}
\bibliography{ref}

\end{document}